\documentclass[prl,aps,twocolumn,showpacs,draft]{revtex4}

\begin{document}

\title{ General Solutions of Braneworlds under the Schwarzschild Ansatz }  
\author{			Keiichi Akama*}
\author{			Takashi Hattori**}
\author{			Hisamitsu Mukaida*}
\affiliation{	*Department of Physics, Saitama Medical University,
 			 Saitama, 350-0495, Japan}
\affiliation{	**Department of Physics, Kanagawa Dental College,
 			 Yokosuka, 238-8580, Japan}
\date{\today}

\begin{abstract}
{\it General solutions} for {\it braneworld dynamics}
	coupled with the bulk Einstein equation are derived
	under the Schwarzschild ansatz. 
They 
	relate the brane metric to the exterior configurations,
and establish fine tuning conditions for the braneworlds 
	to reproduce successes of the Einstein gravity. 
\end{abstract}




\maketitle

Einstein gravity is successful in 
	(i) explaining the Newton's law of universal gravitation 
	for moderate distances, and in
	(ii) predicting the observed light deflections and 
	planetary perihelion precessions due to stellar gravity.
They are derived via the Schwarzschild solution 
	of the Einstein equation
	based on the ansatze, 
	(a) staticity, 
	(b) spherical symmetry,
	(c) asymptotic flatness, and 
	(d) emptiness except for the core of the sphere.

There are many plausible reasons that tempt us
	to take our 3+1 dimensional curved spacetime 
	as a ``braneworld" embedded in higher dimensions 
	\cite{Fronsdal}--\cite{Gregory:2008rf}.
The braneworld theories would be expected to inherit the successes
	of the Einstein gravity. 
It is, however, not true because 
	there exists no Einstein equation of the braneworld itself.
The brane metric alone is insufficient to specify 
	the brane configuration in the bulk. 
Instead, we should use the brane position variable,
	which obeys its own dynamics, 
	but not the Einstein equation \cite{RegTei}. 
To reproduce the Einstein gravity on the brane, 
	we need finely to tune exterior configurations in general. 
A notable exception is the ``brane induced gravity"
	\cite {Akama82}, \cite {Akama87}, \cite{AH},
	\cite {BIG}, \cite {DvaliGabadadze}, \cite {Akama06}, 
	\cite{Gabadadze:2007dv},
	where the brane Einstein equation is supplied 
	by quantum ``induced gravity" \cite{InducedGravity}
	with a composite metric.
Otherwise, however, closer examinations are desiered.
Any entity can be physical only through its dynamics.
Hence, we derive {\it general solutions} for {\it braneworld dynamics}
	coupled with the bulk Einstein equation 
	under the {Schwarzschild ansatz} (a)--(d), 
	and establish (fine tuning) conditions to reproduce 
	the successful results of the Einstein gravity. 
We first illustrate them in (A) Nambu-Goto brane 
	in asymptotically flat bulk, 
	and extend it to (B) warped bulk, 
	and to (C) Einstein-Hilbert brane.

Let us consider a 3+1 dimensional braneworld 
	embedded in 4+1 dimensions. 
Let $X^I$ be the bulk coordinate, and
	$\hat g_{IJ}(X^K)$ be the bulk metric 
	at the point $X^K$ \cite{notation}.
Let the brane be located at $X^I=Y^I(x^\mu)$ in the bulk, 
	where $x^\mu$ are parameters.
The dynamical variables of the system are 
	$Y^I(x^\mu)$, $\hat g_{IJ}(X^K)$ and matter fields, 
	but not the brane metric 
	$g_{\mu\nu}=Y^I_{,\mu} Y^J_{,\nu}\hat g _{IJ}(Y^K)$
	\cite{notation}.
Here we consider the simplest models given by the action,
\begin{eqnarray}
	-\!\int\!\! \sqrt{- g } (\lambda + \gamma R) d^4 x
	-\int\!\!\sqrt{-\hat g}(\hat\lambda+\hat\gamma\hat R)d^5 X
	+ S_{\rm m},
  \label{action}
\end{eqnarray}
where $g=\det g_{\mu\nu}$, $\hat g =\det \hat g _{IJ}$, 
	$ R=R^\mu_{\ \mu}$, $\hat R =\hat R^I_{\ I}$,
	$ R_{\mu\nu}=R^\lambda_{\ \mu\nu\lambda}$, 
	$\hat R_{IJ}=\hat R^K_{\ IJK}$, 
	$ R^\rho_{\ \mu\nu\lambda }$ and $\hat R^L_{\ IJK}$ 
	are the brane and bulk curvature tensors
	written in terms of $g_{\mu\nu}$ and $\hat g _{IJ}$,
 	respectively,
	$ S_{\rm m}$ is the matter action,
	and  $ \lambda $, $ \gamma$, $\hat \lambda$, and $\hat \gamma$ 
	are constants.
We add no artificial fine-tuning terms.
The equations of motion derived from (\ref{action}) are %
\begin{eqnarray}&&
	\left[- \lambda g^{\mu\nu}
	+\gamma \left(R^{\mu\nu}-\frac{1}{2} R g^{\mu\nu}\right)
	- T^{\mu\nu}\right]
	Y^I_{;\mu\nu}=0,\ \ \ \ \ \ 
  \label{NGRT}
\\&&\ \ 
	-\hat \lambda \hat g _{IJ} 
	+\hat \gamma \left(\hat R_{IJ}-\frac{1}{2}\hat R \hat g _{IJ}\right)
	= \hat T_{IJ},
  \label{BE}
\end{eqnarray}
and those for matters,
where $T_{\mu\nu}$ and $\hat T_{IJ}$ are the energy momentum tensors 
	with respect to $g_{\mu\nu}$ and $\hat g_{IJ}$, 
	respectively \cite{notation}.
Eq.\ (\ref{NGRT}) is 
	the Nambu-Goto+Regge-Teitelboim equation \cite{RegTei}
	for the {\it braneworld dynamics}, and
	eq.\ (\ref{BE}) is the bulk Einstein equation. 
The brane Einstein equation does not hold.
Hence, we solve (\ref{NGRT}) and (\ref{BE}) at the brane, 
	and examine their implications.
(If necessary, they can be continued into the bulk in known methods
	\cite{GarrigaTanaka}--\cite{Creek:2006je}, \cite{Gregory:2008rf}.)
We first illustrate them in a simple model in (A),
	and discuss extensions in (B) and (C).

\noindent
(A) {\it Nambu-Goto Brane in Asymptotically Flat Bulk} 
($\hat \gamma \ne 0$, $\lambda\ne0$, $\hat \lambda= \gamma=0$) \ \ 
We introduce the Gaussian normal coordinate 
	$(X^\mu, X^4)=(x^\mu, \xi) $ with $\xi=0$ at the brane,
and expand the bulk metric $\hat g_{IJ}$ in $\xi$ as
\begin{eqnarray}&&
	\hat g _{\mu\nu} 
	= g _{\mu\nu}-2\xi K _{\mu\nu}
	+\xi^2 C_{\mu\nu}+O(\xi^3),\ \ \ \ \nonumber
  \\ &&
	\hat g _{\mu4}=O(\xi^3), 
   \ \ \ \ \ \ 
	\hat g _{44}=-1+ O(\xi^3),
  \label{ghat=}
\end{eqnarray}
where $K_{\mu\nu}$ and $C_{\mu\nu}$ are functions of $x^\mu$.
The former is the extrinsic curvature: 
	$K_{\mu\nu}=Y^4_{;\mu\nu}=\hat\Gamma^4_{\mu\nu}|_{\xi=0}$
	with the bulk affine connection $\hat\Gamma^I_{JK}$.
We introduce the time, radial, polar and azimuth coordinates,
	$x^0=t$, $x^1=r$, $x^2=\theta$, and $x^3=\phi$, respectively,
	according to ansatze (a) and (b),   
	which imply that $\hat g_{IJ}$ is diagonal, and
	$\hat g_{00}$, $ \hat g_{11}$, 
	$\hat g_{22}=\hat g_{33}/\sin^2\theta$, and $ \hat g_{44}$ 
	are functions of $r$ and $\xi$. 
We parameterize them with
\begin{eqnarray}&&
	g_{\mu\nu}
	={\rm diag}(f, h, r^2, r^2\sin^2\theta),
\label{gmunu}
\\&&
	K^\mu _\nu
	={\rm diag}(a, b, c, c),\ \ 
	U^\mu _\nu 	={\rm diag}(u, v, w, w),\ \ \ \  
  \label{uvw}
\end{eqnarray}
where 	$ U^\mu _\nu 
	=(\hat R^{4\mu}{}_{\nu4}
	-\delta^\mu_\nu\hat R^{4\lambda}{}_{\lambda 4})|_{\xi=0}$
	represents the degrees of freedom of 
	$C_{\mu\nu} = K_\mu^\lambda K_{\lambda\nu}
	+\hat R^{4 }{}_{\mu \nu4}|_{\xi=0}$,
and
	$f$, $h$, $a$, $b$, $c$, $u$, $v$, and $w$ are functions of $r$.
Then, eq.\ (\ref{NGRT}) implies 
\begin{eqnarray}
	a+b+2c=0,
  \label{NGRTa} 
\end{eqnarray}
while eq.\ (\ref{BE}) at the brane leads to
\begin{eqnarray}&&
	-(h^{-1})'/r+(1-h^{-1})/r^2-2bc-c^2+u=0, \ \ \ \ \ 
\label{BE00}
\\&&%
	-f'/rfh+(1-h^{-1})/r^2-2ac-c^2+v=0,
  \label{BE11}
\\&&
	- ((fh)^{-1/2}f')'/2(fh)^{ 1/2}-(f/h)'/2rf \nonumber
\\ &&\hskip92pt %
	-ab-ac-bc+w=0, \label{BE22}
\\ &&%
	- ((fh)^{-1/2}f')'/2(fh)^{ 1/2}-(f/h)'/rf \nonumber 
\\&&\hskip13pt 
	+(1-h^{-1})/r^2-ab-2ac-2bc-c^2=0, \label{BE44}
\end{eqnarray}
where ansatz (d) is used, and
	contributions from the brane action to $\hat T_{IJ}$ are neglected.
We have five equations (\ref{NGRTa})--(\ref{BE44})
	for eight variables $f$, $h$, $a$, $b$, $c$, $u$, $v$, and $w$.
Accordingly, the general solution involves three arbitrary functions,
	for which we here choose $a$, $c$ and $v$.

We first eliminate $f$, $f'$, $f''$ and $b$
	from (\ref{NGRTa}), (\ref{BE11}) and (\ref{BE44})
	to obtain the differential equation, 
\begin{eqnarray}
	(rh^{-1})'
	&=&1+r^2V+\frac{ 2r^2(4V+rV'-A)}{h(1+r^2V)-3} \label{deh}\ \ \ \ 
\\
{\rm with}\ \ 
	V&=&v-2ac-c^2, \ 
\\\ \ \ \ \ \ \ 
	A &=& 2a^2 + 4ac + 6c^2, \label{Xi=}
\end{eqnarray}
To solve eq.\ (\ref{deh}), we rewrite it into the form
\begin{eqnarray}&&\hskip-30pt
	Z'=P+Q/(1+Z/r)      \label{key}\\&&
	\hskip-27pt {\rm with}\ \ \ 
	Z= 3(h^{-1}-1)/2-Vr^2/2,   \label{zeta=}\\&&
	P= (A-4V-3V'r/2) r^2,   \label{P=}\\&& 
	Q= (1+Vr^2)(A -4V- V'r) r^2/2.   \label{Q=}
\end{eqnarray}
A sufficient condition for existence of the unique
	solution to (\ref{key}) 
	with the initial value $Z=Z_{(0)}$ at $\rho\equiv1/r=0$ 
	is that, in a region $D$ in the $\rho$-$Z$ plane
	including $(0,Z_{(0)})$,
\begin{eqnarray}&&
	P / \rho^2+Q/ \rho^2 (1+Z\rho) {\rm \ is\ continuous,\ and}
 \label{cont}\ \ \ \ \ \ 
\\&&
	\exists M:{\rm real}\ \ 
	\forall (\rho,z_1), (\rho,z_2)\in D  \nonumber 
\\&&\ \ \ \ 
	|Q/\rho(1+z_1\rho)(1+z_2\rho) | \le M, \ \ \ \ 
 \label{Lip}
\end{eqnarray}
which we assume here.
Then, the solution to (\ref{key}) is given by
	$ Z=\lim_{n\rightarrow\infty} Z_{(n)}$ with
\begin{eqnarray}
	Z_{(n)}=Z_{(0)}
	+\int_{\infty}^r \left[P +\frac{Qr}{r+Z_{(n-1)}}\right]dr.
	\  (n\ge1)
 \label{Z_(n)}
\end{eqnarray}
Another convenient form of the solution would be given by
	series expansion in $1/r$.
Eqs.\ (\ref{cont}) and (\ref{Lip}) indicate that
	$P r=-Qr=$finite at $r=\infty$.
Hence, we assume existence of (at least, 
	asymptotic) expansions:
\begin{eqnarray}
	Z&=& Z_0+Z_1/r+ Z_2/r^2+\cdots,  \label{zeta=...}\\
	P&=& P_1/r+P_2/r^2+\cdots,  \label{P=...}\\
	Q&=& Q_1/r+Q_2/r^2+\cdots  \label{Q=...},
\end{eqnarray}
where 
	$Z_i$, $P_i$ and $Q_i$ ($i=1,2,\cdots$) are constants, and $Z_0=Z_{(0)}$.
Then, eq.\ (\ref{key}) implies the recurrence,
\begin{eqnarray}
	Z_n=
	-\sum_{k=0}^{n-1} 
	\frac{ Z_k [(n-1) Z_{n-k-1}+2 P_{n-k}]}{2n}
	-\frac{\Sigma_{n+1}}{n},\ \ \ 
\label{zeta_n=}
\end{eqnarray}
for $n\ge1$, where $\Sigma_n=P_n+Q_n $, or, explicitly,
\begin{eqnarray}&&
	Z_1=-Z_0 P_1-\Sigma_2,
\\ &&
	Z_2=[ Z_0^2 P_1- Z_0 (P_1^2-Q_2)
	+P_1 \Sigma _2- \Sigma _3]/2, \ \ \ 
\\&& 
	Z_3=-[
	2 Z_0{}^3 P_1
	+ Z_0{}^2(5P_1^2+2Q_2) \nonumber
\\&&\ \ \ \ \ \ \ \ 
	+ Z_0 (5P_1^3+4P_1P_2
	+7P_1Q_2-2Q_3) \nonumber
\\&&\ \ \ \ \ \ \ \ 
	+P_1{}^2 \Sigma _2- P_1\Sigma _3+ 2Q_2\Sigma _2
	+2\Sigma _4]/6,
\\&& 
	\ \ \  \cdots\ \ \ \ \cdots \ \ \ \ \cdots. \nonumber
\end{eqnarray}
Once the function $Z$ of $r$ is thus determined, 
	{\it the general solution} of the braneworld
	is given by
\begin{eqnarray}
	h &=&1/(1+2Z/3r+Vr^2/3),\label{sol h}\\
	f &=&\exp\int_\infty^r\frac{h(1+Vr^2)-1}{r}dr, \label{sol f}\\
	u &=& [{(r/h)'-1}]/{r^2} -2ac-3c^2, \label{sol u}\\
	w &=&-(u+v)/2, \label{sol w}\\
	b &=&-a-2c, \label{sol b}
\end{eqnarray}
	with three arbitrary functions $a$, $c$ and $v$.
The braneworld dynamics (\ref{NGRT}) is crucially important.
If we omitted it, and used only the bulk Einstein equation (\ref{BE}),
	we would have had the wrong solution without (\ref{sol b}) and with
\begin{eqnarray}
	A&=& -2ab-4ac-4bc-2c^2, \label{Rac}\\
	u &=& [{(r/h)'-1}]/{r^2} +2bc+c^2, \label{sol u'}
\end{eqnarray}
in the places of (\ref{Xi=}) and (\ref{sol u}), respectively,
	which would cause large differences in the consequences below.

Now we compare our general solution 
	with the results (i) and (ii) of the Einstein gravity
	in 3+1 dimensions.
The latter is given by (\ref{BE00}) and (\ref{BE11}) with $a=b=c=u=v=0$, and 
	is solved to give the Schwarzschild solution $ f=h^{-1}=1-\mu/r $
with an arbitrary constant $\mu$, 
	which is determined phenomenologically.
In geodesic motions of objects, 
	the function $V_{\rm N}\equiv f-1=-\mu/r $ 
	serves as the Newtonian potential for gravitational forces
	for moderate distances,
	as renders a precise explanation 
	for the Newton's law of the universal gravitation. 
The angles $\Delta \varphi_{\rm E}$ and $\Delta \psi_{\rm E}$
	 of light deflection and planetary perihelion precession, respectively, 
	due to the stellar gravity are given by
	$\Delta \varphi_{\rm E} = 2\mu/r_0$ and
	$\Delta \psi_{\rm E} = 3\pi\mu/(1-e^2)a$, 
where $r_0$ is the minimum of $r$ on the light orbit, and 
	$e$ and $a$ are, respectively, the eccentricity and the average distance 
	of the planetary orbit.
These predictions precisely match with the observations.

Now we turn to our general solution. 
The Newtonian potential $V_{\rm N}\equiv f-1$ 
	has large arbitrariness owing to those of $a$, $c$ and $v$.
In fact, we can always choose appropriate $a$, $b$, $c$, $u$, $w$ and $v$ 
	for arbitrary $f$ and $h$ with $R\ge0$, 
	while we have no solution for $R<0$.
Therefore, we cannot predict that $V_{\rm N}=-\mu/r$,
	unlike the Einstein gravity. 
The singularity structures also depend on exterior configurations. 
The conditions (\ref{cont}) and (\ref{Lip}) 
	would be useful in investigating them.

In ordinary observations with $r>>\mu$,
	the $\mu/r$ term dominates $V_{\rm N}$.
Then, we parametrize the deviations 
	from the results of the Einstein gravity by
\begin{eqnarray}
	f &=& 1-\mu/r+f_2/r^2 +f_3/r^3+\cdots, \label{f=...}\\
	h^{-1} &=& 1-\mu/r +\varepsilon_1 /r
	+\varepsilon_2 /r^2+\cdots \label{h-1=...}
\end{eqnarray}
with constants $f_i$ ($i\ge2 $) and $\varepsilon_i$ ($i\ge1$).
The asymtotic flatness implies 
	$f,h\rightarrow 1$ and $f',h'\rightarrow 0$
	as $r \rightarrow \infty$.
Then, eq.\ (\ref{BE11}) (with $\hat \lambda=0$) implies
	 $Vr^2\rightarrow 0$.
This and eqs.\ (\ref{P=})--(\ref{Lip}) indicate
	$Ar^2\rightarrow0$,
	and therefore (\ref{Rac}) impies $ar, br, cr \rightarrow0$.
These and $ V r^2\rightarrow0$ indicate $vr^2\rightarrow0$.
Hence, we assume existence of (at least, asymptotic) expansions 
	of the arbitrary functions:
\begin{eqnarray}
	a &=& a_2/r^2+ a_3/r^3+\cdots,  \label{a=...}\\
	c &=& c_2/r^2+ c_3/r^3+\cdots,  \label{b=...}\\
	v &=& v_3/r^3+ v_4/r^4+\cdots  \label{v=...}
\end{eqnarray}
with constants $a_i$, $c_i$ ($i\ge2 $) and $v_i$ ($i\ge3$).
Then, the present general solution indicates 
\begin{eqnarray}
&&\hskip-5mm	
	\varepsilon_1 = v_3, \\
&&\hskip-5mm	
	f_2 = 
	3\mu v_3/4 -v_4 -a_2{}^2-2{c_2}^2, \\
&&\hskip-5mm
	\varepsilon_2 = 
	\mu v_3/2 -v_4 -2a_2{}^2-2a_2c_2-5{c_2}^2, \\
	&&\cdots\ \ \ \ \cdots\ \ \ \ \cdots\nonumber 
\end{eqnarray}

In terms of them,
	the angles $\Delta \varphi$ and $\Delta \psi$
	of light deflection and planetary perihelion precession,
 	respectively, due to the stellar gravity are given by,
	up to $O(r^3)$,
\begin{eqnarray}
	\Delta \varphi 
	&\approx& \Delta \varphi_{\rm E} (1-\varepsilon_1/2\mu)
	=\Delta \varphi_{\rm E}(1-v_3/2\mu),
\\
	\Delta \psi &\approx& \Delta \psi_{\rm E}
	[1-(\mu\varepsilon_1+2f_2)/3\mu^2]
\cr 
	&=& \Delta \psi_{\rm E}
	\left[1-\frac{5v_3}{6\mu}
	+\frac{2(v_4+{a_2}^2+2{c_2}^2)}{3\mu^2}
	\right].\ \ \ 
\label{DpsiE}
\end{eqnarray}
Consequently, the condition to match the observations is
\begin{eqnarray}
	v_3\approx0,\ \ \ \ v_4+a_2{}^2+2{c_2}^2\approx 0. 
\label{cond}
\end{eqnarray}
The condition for the bulk Einstein equation 
	(\ref{BE00}--\ref{BE44}) at the brane
	to imply the brane-Einstein equation would be 
\begin{eqnarray}&&
	u=2bc+c^2,\ \ 
	v=2ac+c^2,\ \  
	w=ab+ac+bc.\ \ \ \ \ \  \label{cond uvw}
\end{eqnarray}
They are, however, not all physical.
Under the dynamics, (\ref{cond uvw}) leads to $ a^2+2ac+3c^2=0$. 
Hence, the brane Einstein equation holds only when
\begin{eqnarray}&&%
	a=b=c=u=v=w=0. \label{abc0}
\end{eqnarray}

\noindent
(B){\it Warped Bulk}
($\hat \lambda \ne 0$, $\hat \gamma \ne 0$, $\lambda\ne0$,
$\gamma=0$) \ \ 
The dynamics (eqs.\ (\ref{NGRT}) and (\ref{BE})) 
	has a solution with a flat brane at $|\xi|\le\delta$
	in empty warped bulk \cite{RS}: 
\begin{eqnarray}
	\hat g_{IJ}={\rm diag}(F,-F,-Fr^2,-F r^2\sin^2\theta,-1),
	\label{RS}
\end{eqnarray}
where $\delta$ is an infinitesimal constant, and
	$F$ is a smooth function of $\xi$ such that $F'|_{\xi=0}=0$
	and $F=e^{-2k|\xi|}$ for $|\xi|\ge\delta$ 
	with $k^2=-\hat\lambda/6\hat\gamma$.
We seek for the general solutions,  
	where the metric $\hat g_{IJ}$
	approaches (\ref{RS}) as $r\rightarrow\infty$.
Hence, $\hat \Gamma^4_{\mu\nu}|_{\xi=\pm\delta}
	\rightarrow\mp kg_{\mu\nu}$,
	$\hat \Gamma^4_{\mu\nu}|_{\xi=0}
	\rightarrow 0$,
	and $\hat R^{4\mu}{}_{\nu4}|_{\xi=\pm\delta}
	\rightarrow k^2\delta^\mu _\nu$.
The ansatz (c) is retained on the brane, but 
	(d) is slightly violated 
	by matter distributed in $|\xi|\le\delta$.
We assume that the brane-generating interactions are
	much stronger than gravity at short distances of $O(\delta)$,
	while their gravitations are much weaker than
	those by the core of the sphere.
Then, $\hat T_{\mu\nu}$ in $|\xi|\le\delta$ 
	becomes independent of $r$, and so does 
	$(\hat \Gamma^4_{\mu\nu}|_{\xi=\pm\delta}
	-\hat \Gamma^4_{\mu\nu}|_{\xi=0})$ \cite{Israel}, 
	so that it is equal to its asymptotic value
	$\mp k g_{\mu\nu}$.
We define the functions 
	$\tilde a$, $\tilde b$, $\tilde c$, 
	$\tilde u^\pm$, $\tilde v^\pm$ and $\tilde w^\pm$ of $r$ by
\begin{eqnarray}&&
	\hat \Gamma^4_{\mu\nu}|_{\xi=0}
	=\hat \Gamma^4_{\mu\nu}|_{\xi=\pm\delta}\pm k g_{\mu\nu}
	={\rm diag}(\tilde a, \tilde b, \tilde c, \tilde c),
  \label{RSabcdef}\\&&
	(\hat R^{4\mu}{}_{\nu4} 
	-\delta^\mu_\nu
	\hat R^{4\lambda}{}_{\lambda 4})|_{\xi=\pm\delta} \nonumber
\\&&\ \ \ \ \ \ \ \ \ \ \ \ 
	=-3k^2 \delta^\mu_\nu +{\rm diag} 
	 (\tilde u^\pm, \tilde v^\pm, \tilde w^\pm, \tilde w^\pm).
	\ \ \ \ \ 
\end{eqnarray}
Then, eq.\ (\ref{NGRT}) with $\gamma=0$ and eq.\ (\ref{BE}) at $\xi=\pm\delta$
	imply the same equations as (\ref{NGRTa})--(\ref{BE44}) in (A)
	with $\tilde a$, $\tilde b$, $\tilde c$, 
	$\tilde u^\pm \mp2k \tilde a$,
	$\tilde v^\pm \mp 2k \tilde b$ and 
	$\tilde w^\pm \mp 2k \tilde c$
	in the places of $a$, $b$, $c$, $u$, $v$ and $w$, respectively. 
Therefore, we obtain the same results as in (A) above,
	though the geometrical interpretations are different.

\noindent
(C) {\it Einstein Hilbert Brane} 
($\hat \gamma \ne 0$, $\gamma\ne0$, $\hat \lambda =\lambda=0$) 
The brane dynamics (\ref{NGRT}) implies, instead of (\ref{NGRTa}) in (A), 
\begin{eqnarray}&&\hskip-20pt
	 [-(h^{-1})'/r+(1-h^{-1})/r^2]a \nonumber
\\&&\hskip-20pt 
	+ [-f'/rfh+(1-h^{-1})/r^2]b \nonumber 
\\&&\hskip-20pt 
	+2 [- ((fh)^{-1/2}f')'/2(fh)^{ 1/2}-(f/h)'/2rf]c=0, 
  \label{RTa} 
\end{eqnarray}
	while eqs.\ (\ref{BE00})--(\ref{BE44}) are unchanged.
If we eliminate $f$, $f'$, $f''$, $b$, $u$, and $w$ 
	from eqs.\ (\ref{BE00})--(\ref{BE44}) and (\ref{RTa}), we obtain
\begin{eqnarray}&&
	(rh^{-1})'=1+r^2V     \nonumber 
\\&&
	+\frac{r^2[rV'(V-2ac-4c^2)+5V^2-2Va(2a+c)]}
	{h(1+r^2V)(V-2ac-4c^2)+2(a+2c)(2a+c)}.\ \ \ \ \ \ \ 
  \label{dehRT}
\end{eqnarray}
Eq.\ (\ref{dehRT}) has the same structure as (\ref{deh})
	as far as $h$ is concerned.
Hence, we can apply the same method as that in (A) above.
In the present case ($\lambda=0$), the condition (\ref{cond uvw})
	automatically implicates the brane dynamics (\ref{RTa}).
Therefore, (\ref{cond uvw}) just gives
	the condition for the brane Einstein equation to hold, 
	but not (\ref{abc0}).

In any cases, the successful results of the Einstein gravity 
	are reproduced through fine tuning,
	and are not ``predicted" by the braneworld theories.
It would be an urgent and important challenge to seek for
	physical foundations of the fine-tuning conditions.

We would like to thank 
Professor T.~Inami, 
Professor I.~Oda, 
Professor G.~R.~Dvali,
Professor G.~Gabadadze,
Professor M.~E.~Shaposhnikov, 
Professor T.~Asaka, 
Professor I.~Antoniadis, 
Professor M.~Giovannini,
Professor S.~Randjbar-Daemi, 
Professor R.~Gregory, 
Professor P.~Kanti, 
Professor G.~W.~Gibbons, 
Professor K.~Hashimoto, 
Professor E.~J.~Copeland, 
Professor D.~L.~Wiltshire, 
Professor I.~P.~Neupane,
Professor R.~R.~Volkas, and 
Professor A.~Kobakhidze 
for invaluable discussions and for their kind hospitalities over us
when we visited them.
This work was supported by Grant-in-Aid for Scientific Research,
No.\ 13640297, 17500601, and 22500819
from Japanese Ministry of Education, Culture, Sports, Science and Technology.


\begin{thebibliography}{99}
\bibitem{Fronsdal}
C.~Fronsdal, Nuovo Cim.\ {\bf 13}, 5 (1959).
\bibitem{Joseph}
D.~W.~Joseph, Phys.\ Rev.\ {\bf 126}, 319 (1962).
\bibitem{RegTei}
T.Regge and C.Teitelboim, 
in {\it Marcel Grossman Meeting on Relativity, 1975}(North Holland, 1977) 77.
\bibitem{Akama82}
K.~Akama, Talk given at International Symposium on Gauge Theory and Gravitation, Nara, Japan, 1982; Lect.\ Notes in Phys.\ {\bf 176}, 267 (1983).
\bibitem{RubShap}
V.~A.~Rubakov and M.\ E.\ Shaposhnikov, Phys.\ Lett.\ {\bf B125}, 136 (1983).
\bibitem{Maia}
M.~D.~Maia, J.\ Math.\ Phys.\ {\bf 25}, 2090 (1984). 
\bibitem{Visser}
M.~Visser, Phys.\ Lett.\ {\bf B159}, 22 (1985).
\bibitem{Pavsic}
M.~Pav\v si\v c, Class.\ Quant.\ Grav.\ {\bf 2}, 869 (1985);
  Phys.\ Lett.\  A {\bf 107}, 66 (1985).
\bibitem{NicolaiWetterich}
H. Nicolai and C. Wetterich. Phys.\ Lett.\ {\bf B150}, 347 (1985).
\bibitem{RandjbarDaemi:1985wg}
  S.~Randjbar-Daemi and C.~Wetterich,
  Phys.\ Lett.\  B {\bf 166}, 65 (1986).
\bibitem{GibbonsWiltshire}
G. W.  Gibbons and D. L. Wiltshire, Nucl.\ Phys.\ {\bf B287}, 717 (1987). 
\bibitem{Akama87}
K.~Akama,  Prog.\ Theor.\ Phys.\ 
{\bf 78}, 184 (1987), {\bf 79}, 1299 (1988), {\bf 80}, 935 (1988).
\bibitem{Nakamura:1989mx}
  A.~Nakamura, S.~Hirenzaki and K.~Shiraishi,
  Nucl.\ Phys.\  B {\bf 339}, 533 (1990).
\bibitem{Antoniadis}
I. Antoniadis, Phys. Lett. {\bf B246}, 377 (1990).
\bibitem{Polchinski}
J.~Polchinski, Phys.\ Rev.\ Lett.\ {\bf 75}, 4724--4727 (1995).
\bibitem{HoravaWitten}
P.~Horava and E.~Witten, 
Nucl.\ Phys.\ {\bf B460}, 506--524 (1996); Nucl.\ Phys.\ {\bf B475}, 94--114 (1996).
\bibitem{ADD}
N.~Arkani-Hamed, S.~Dimopoulos, and G.~Dvali, 
Phys.\ Lett.\ {\bf B429}, 263-272 (1998); 
Phys.\ Rev.\ {\bf D59}, 086004 (1999).
\bibitem{AADD}
I.~Antoniadis, N.~Arkani-Hamed, S.~Dimopoulos, and G.~Dvali, 
Phys.\ Lett.\ {\bf B436}, 257-263 (1998).
\bibitem{RS}
L.~Randall, R.~Sundrum, 
Phys.\ Rev.\ Lett.\ {\bf 83}, 3370-3373 (1999); 4690-4693 (1999).
\bibitem{AH} 
K.~Akama and T.\ Hattori, Mod.\ Phys.\ Lett.\ {\bf A15}, 2017 (2000).
\bibitem{BIG} 
G.\ Dvali, G.\ Gabadadze, and M.\ Porrati, 
Phys.\ Lett.\ {\bf B485}, 208 (2000).
\bibitem{SMS}
T.\ Shiromizu, K.\ I.\ Maeda, and M.\ Sasaki, Phys.\ Rev.\ {\bf D62}, 024012 (2000). 
\bibitem{DvaliGabadadze}
G.~Dvali and G.~Gabadadze,
Phys.\ Rev.\ D {\bf 63}, 065007 (2001).
\bibitem{GarrigaTanaka}
J.~Garriga and T.~Tanaka, Phys.\ Rev.\ Lett.\ {\bf 84}, 2778 (2000).
\bibitem{Visser:2002vg}
  M.~Visser and D.~L.~Wiltshire,
  Phys.\ Rev.\  D {\bf 67}, 104004 (2003)
  [arXiv:hep-th/0212333].
\bibitem{Casadio:2002uv}
  R.~Casadio and L.~Mazzacurati,
  Mod.\ Phys.\ Lett.\  A {\bf 18}, 651 (2003)
  [arXiv:gr-qc/0205129].
\bibitem{Bronnikov:2003gx}
  K.~A.~Bronnikov, V.~N.~Melnikov and H.~Dehnen,
  Phys.\ Rev.\  D {\bf 68}, 024025 (2003)
  [arXiv:gr-qc/0304068].
\bibitem{Kanti}
  P.~Kanti,
  Int.\ J.\ Mod.\ Phys.\  A {\bf 19}, 4899 (2004);
  J.\ Phys.\ Conf.\ Ser.\  {\bf 189}, 012020 (2009).
\bibitem{Creek:2006je}
  S.~Creek, R.~Gregory, P.~Kanti and B.~Mistry,
  Class.\ Quant.\ Grav.\  {\bf 23}, 6633 (2006)
  [arXiv:hep-th/0606006].

\bibitem{Akama06}
  K.~Akama, arXiv: gr-qc/0607106 (2006);
  K.~Akama and T.~Hattori, submitted for publication (2009).
\bibitem{Gabadadze:2007dv}
  G.~Gabadadze,
  Nucl.\ Phys.\ Proc.\ Suppl.\  {\bf 171}, 88 (2007).
\bibitem{DavidsonGeorgeKobakhidzeVolkasWali}
A.~Davidson, D.~P.~George, A.~Kobakhidze, R.~R.~Volkas, K.~C.~Wali, 
Phys.\ Rev.\ {\bf D77}, 085031 (2008).


\bibitem{Wetterich08}
  C.~Wetterich,
  Phys.\ Rev.\  D {\bf 78}, 043503 (2008);
   {\bf 81}, 103508 (2010);
  Phys.\ Rev.\ Lett.\  {\bf 102}, 141303 (2009).


\bibitem{Kodama} 
  Y.~Kodama, K.~Kokubu and N.~Sawado,
  Phys.\ Rev.\  D {\bf 78}, 045001 (2008);
  Phys.\ Rev.\  D {\bf 79}, 065024 (2009).
\bibitem{Gregory:2008rf}
  R.~Gregory,
  Lect.\ Notes Phys.\  {\bf 769}, 259 (2009).
\bibitem{InducedGravity}
A.~D.~Sakharov,  Dokl.\ Akad.\ Nauk SSSR {\bf 177}, 70 (1967)
	[{Sov.\ Phys.\ Dokl.} {\bf 12}, 1040 (1968)];
K.~Akama, Y.~Chikashige and T.~Matsuki,
Prog.\ Theor.\ Phys.\  {\bf 59}, 653 (1978);
K.~Akama, Y.~Chikashige, T.~Matsuki and H.~Terazawa,
	 {Prog.\ Theor.\ Phys.} {\bf 60}, 868 (1978);
K.~Akama,  {Prog.\ Theor.\ Phys.} {\bf 60}, 1900 (1978);
A.~Zee, Phys.\ Rev.\ Lett.\ {\bf 42}, 417 (1979);
S.~L.~Adler, Phys.\ Rev.\ Lett.\ {\bf 44}, 1567 (1980); 
  C.~Barcelo, S.~Liberati and M.~Visser,
  Living Rev.\ Rel.\  {\bf 8}, 12 (2005);
B.~Broda and M.~Szanecki,
  Phys.\ Lett.\  B {\bf 674}, 64 (2009).





\bibitem{notation}
{Throughout this paper, uppercase suffices run 0 to 4, 
	lowercase ones, 0 to 3, 
	the suffices following comma indicate differentiations
	with respect to the coordinate variables,
	and those following semicolon, covariant differentiations.} 
\bibitem{Israel}
  W.~Israel,
  Nuovo Cim.\  B {\bf 44S10}, 1 (1966)
  [Erratum-ibid.\  B {\bf 48}, 463 (1967\ NUCIA,B44,1.1966)].
\end{thebibliography}
\end{document}